\documentstyle[ epsfig,graphicx]{elsart}

\newcommand{\beq}{\begin{equation}}
\newcommand{\eeq}{\end{equation}}
\newcommand{\beqa}{\begin{eqnarray}}
\newcommand{\eeqa}{\end{eqnarray}}

\renewcommand{\vec}[1]{{\bf #1}}

\newcommand{\br}{{\bf r}}

\begin{document}
\begin{frontmatter}
\title{Doping induced inhomogeneity in high-$T_c$ 
superconductors\thanksref{talk}}
\thanks[talk]{Expanded version of a talk presented at the 
ISS 2000, Oct 14-16, Tokyo, Japan}

\author{Ivar Martin  and Alexander V. Balatsky\thanksref{email}}
\thanks[email]{ivar@viking.lanl.gov, avb@lanl.gov}
 
\address{Theoretical Division, Los Alamos National Laboratory, 
	Los Alamos, NM 87545, USA}

\begin{abstract}
Doping and disorder are inseparable in the superconducting cuprates.  
Assuming the simplest possible    disordered doping, we 
construct a semiphenomenological model and analyze its experimental 
consequences. Among the affected experimental quantities are the ARPES spectra 
and thermodynamic properties.
From our model we make a prediction for the width of the local 
superconducting gap distribution with the only experimentally unknown parameter 
being the superconducting correlation length.  Thus, our model provides 
a {\em direct} way of determining the superconducting correlation 
length from a known  experimental gap distribution.
\end{abstract}

\end{frontmatter}


\date{\today}




\section{Introduction}

Ground state of all cuprate superconductors is a Mott insulator at half 
filling. To render this state a superconductor one has to  dope it with 
carriers (holes for p-type superconductors such as LSCO and Bi2212, and   
electrons for n-type, e.g NdSCO). Below we will focus on hole doped 
superconductors, although discussion about intrinsic disorder is relevant for 
electron doped materials as well. In the low doping limit, dopants, such as Sr 
in LSCO and oxygen in Bi2212 are certainly randomly distributed across the 
sample without forming any regular array. The central point we argue for in 
this paper is that {\em effect of carriers and intrinsic disorder produced by 
the very same doping are  inseparable in underdoped and possibly overdoped 
cuprates}.

The microscopic inhomogeneity of the superconducting state state has 
important consequences for the average properties obtained in the bulk 
measurements. Immediate conclusion we draw  is  that even with the perfectly 
grown crystals doping will produce  artificial impurities that disorder the 
system. The affected properties include the average tunneling 
spectra, as the ones obtained by ARPES and  measured specific heat.   Another 
implication is that since the state is inhomogeneous, there should be no 
well defined superconducting transition temperature, much like in the 
conventional granular superconductors \cite{goldman}.  This 
``granularity'' should also clearly enhance the electric and the thermal 
transport 
above $T_c$.

\section{Density of States  and gap distribution in the presence of doping 
induced disorder}

\begin{figure}[htbp]
\begin{center}
\includegraphics[width = 3.0 in]{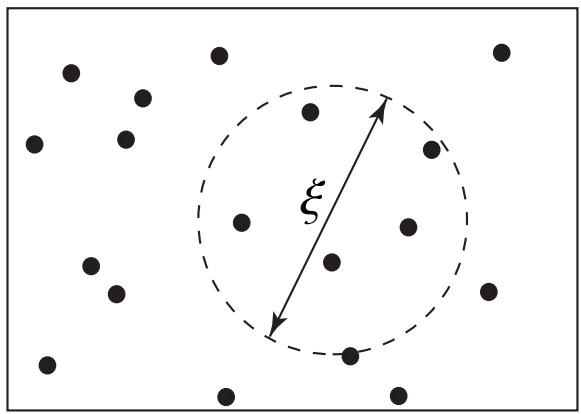}

\vspace{0.5cm}

\caption{The effective local doping at a particular point is defined 
through the random number of dopants which happen to be in the 
$\xi$-vicinity of this point.}
\label{fig:xi}
 \end{center}
\end{figure}

To analyze the effects of doping-induced disorder, we use a minimal 
model of randomly distributed dopants in the reservoir layer. The 
superconducting order parameter can only change on the length scale of the 
superconducting coherence length, $\xi$.  Hence we can coarse-grain the 
Cu-O plane into patches of the size $\xi$ and on these patches define the 
effective dopant concentration as
\beq
n_\xi(\br) = \frac{N_\xi(\br)}{\xi^2},
\eeq
where $N_\xi(\br)$ is the number of dopants in the adjacent reservoir 
plane which happen to be within a circle of radius $\xi$ around the 
point $\br$ (Figure \ref{fig:xi}).  

\begin{figure}[htbp]
\begin{center}
\includegraphics[width = 3.0 in]{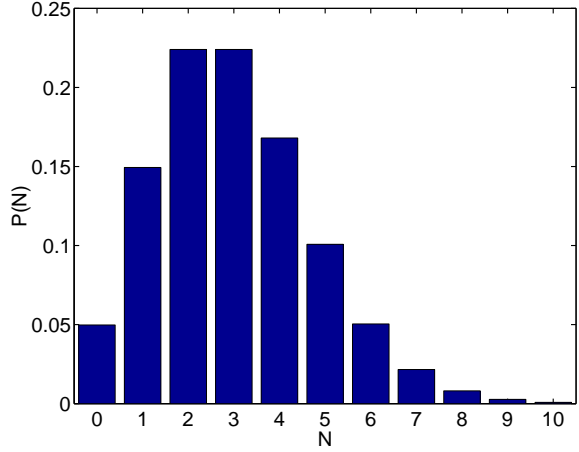}
\caption{Poisson distribution with the average equal to 3.  The distribution is 
defined only on non-negative integers.}
\label{fig:pois3}
\end{center}
\end{figure}

Assuming that the positions of the 
dopants are uncorrelated, which is expected to hold particularly at low 
doping levels, the numbers $N_\xi$ obey the Poisson distribution, with a 
probability density,
\beq\label{PN}
P(N) = \frac{(\bar{n}\xi^2)^N}{N!} \exp(-\bar{n}\xi^2).
\eeq
Here, the average dopant density, $\bar{n}$, is a function of the 
nominal doping fraction, $ \bar{x} = a b \bar{n}$, with $a$ and $b$ the 
in-plane lattice constant.  An example of Poisson distribution is shown in 
Figure \ref{fig:pois3}.

The main characteristic of the Poisson distribution is that its 
dispersion is equal to its average value.
In the limit of large average, $\bar{N} = \bar{n} \xi^2 \gg 1$, the 
Poisson distribution reduces to a Gaussian.  In this limit, the 
probability distribution of the effective local doping fractions becomes
\beq
P(x) = \frac{\xi}{\sqrt{2\pi \bar{x} a b}} 
\exp\left[ \frac{(x - \bar{x})^2}{2 \bar{x}}\frac{\xi^2}{ab}\right].
\label{eq:Px}
\eeq
The distribution of the local doping should lead to a corresponding 
distribution of the gap values over the regions of size $\xi$.  Assuming 
that the distributions of doping is sufficiently narrow, the value of 
the local gap as a function of the local doping can be linearized as
\beq
\Delta(x) = \Delta_0 - \Delta_1 x.
\label{eq:Dx}
\eeq
To the first order approximation, the parameters $\Delta_0$ and 
$\Delta_1$ can be extracted, e.g. from the ARPES data which determines the 
{\em average} gap values as a function of the {\em average} doping. Then, from 
Equations (\ref{eq:Px}) and (\ref{eq:Dx}), the probability
distribution for the {\em local} superconducting gaps follows:
\beq
P(\Delta) = \frac{\xi}{\Delta_1\sqrt{2\pi \bar{x} a b}} 
\exp\left[ \frac{(\Delta - \bar{\Delta})^2}{2\bar{x}\Delta_1^2}
\frac{\xi^2}{ab}\right].
\eeq
The probability distribution of the local values of the gap 
$\Delta(x)$ is completely specified up to the unknown parameter $\xi$.  
Hence, one of the applications of the local gap statistic, and in 
particular the distribution width is the  determination of the 
doping-dependent correlation length, $\xi$.  


To illustrate the smoothing effect of the gap fluctuations on the bulk 
properties we compute the average density of states (DOS), as measured 
by photoemission.  In a conventional $d$-wave BCS superconductor with 
the amplitude of the superconducting gap $\Delta$, the density of state 
is given by 
\beq
N_{\Delta}(\omega) = \left\{ \begin{array}{ll}
N_0 ({\omega}/{\Delta}) K({\omega}/{\Delta}) 
,\quad &{\rm if} \quad \omega <\Delta \\
N_0 K({\Delta}/{\omega})
,\quad &{\rm if} \quad \omega >\Delta
\end{array}\right.
\eeq
Here, $N_0$ is the electronic DOS at the Fermi level in the normal 
state and $K(x)$ is the complete elliptic integral of the first kind.

To determine the average DOS is the presence of the gap amplitude 
fluctuations, we compute the integral 
\beq\label{Nav}
\bar{N}(\omega) = \int{P(\Delta) N_\Delta(\omega) d\Delta}.
\eeq
The average DOS, $\bar{N}(\omega)$, clearly preserves the some rule $\int 
d\omega \bar{N}(\omega) = 1$ due to normalization of $P(\Delta)$.

\begin{figure}[htbp]
\begin{center}
\includegraphics[width = 4.0 in]{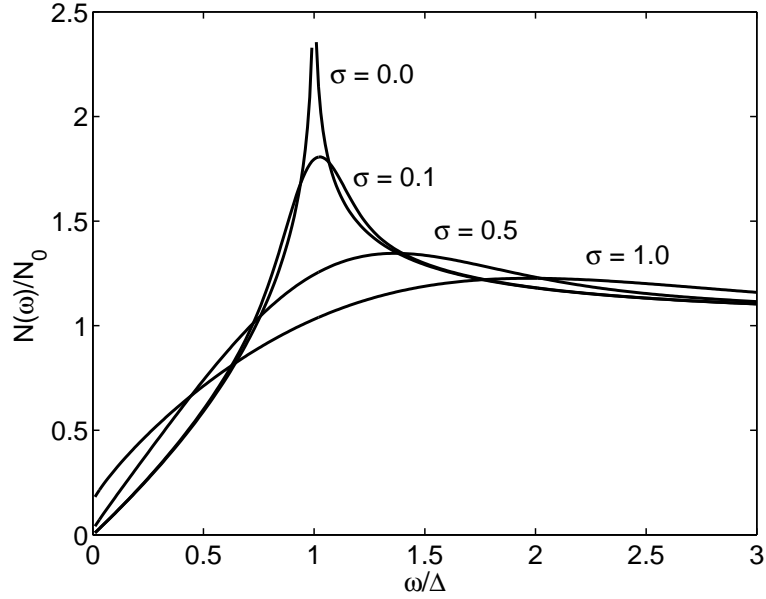}
\vspace{0.5cm}
 
 
\caption{The average BCS $d$-wave density of states (DOS) for different widths 
of the gap distribution, $\sigma$.  Such spatial averaging may explain 
the smooth features commonly observed in the photoemission of optimally 
and underdoped cuprates.  The finite DOS at the Fermi level for large $\sigma$ 
is caused by the zero-gap (normal) regions.}
\label{fig:DOS}
\end{center}
\end{figure}

In Figure \ref{fig:DOS} we show the DOS curves for various widths of 
the gap distribution.  In the ``clean'' case one indeed observes a 
divergence in DOS at the energy equal to the amplitude of the gap.  For a 
finite-width gap distributions, the divergence is replaced by a hump near  
the average gap value.  For wide gap distributions, such that the distribution 
width is comparable to the magnitude of the gap itself (case $\sigma = 1$ in 
Fig. \ref{fig:DOS}), normal regions can be expected to appear in the sample.    
Such regions   give a finite contribution to the DOS at the Fermi level.  This 
contribution is similar in spirit to the well-known effect of non-magnetic 
impurities on the DOS in superconductors \cite{gorkov,lifshitz}.

For the optimally-doped Bi2212, the observed  width of the gap distribution is 
about 10\% of the average\cite{Panaps}.  As is clear from the Figure 
\ref{fig:DOS}, such width corresponds to a rather smooth peak, much like the 
one observed in ARPES \cite{johnson}.

\section{Acknowledgments}
We would like to thank J.C. Davis and S.H. Pan  for useful discussions.  This 
work was 
supported 
by the U.S. DOE.



\end{document}